\documentclass[aps,reprint,superscriptaddress]{revtex4-1}
\usepackage[usenames,dvipsnames]{xcolor}
\usepackage{graphicx}
\usepackage{amsmath}
\usepackage{amssymb}
\usepackage{esint}
\usepackage{amsfonts}
\usepackage{natbib}
\usepackage{psfrag}
\usepackage{wasysym}
\usepackage{float}
\usepackage{epstopdf}
\usepackage{color}
\usepackage[normalem]{ulem}
\usepackage{textcomp}
\usepackage{bm}

\definecolor{cream}{RGB}{222,217,201}

\newcommand{\id}{\mathrm{d}} 

\newcommand{\e}{\mathrm{e}}

\newcommand{\Lag}{\mathcal{L}}

\newcommand{\pd}[2]{\frac{\partial #1}{\partial #2}} 
 
\begin{document}

\title{Elastic Rayleigh-Plateau instability: Dynamical selection of nonlinear states}

\author{Anupam Pandey}
\email[]{ap2338@cornell.edu}
\affiliation{Physics of Fluids Group, Faculty of Science and Technology, University of Twente, P.O. Box 217, 7500AE Enschede, Netherlands}
\affiliation{Department of Biological and Environmental Engineering, Cornell University, Ithaca, NY 14853, USA}
\author{Minkush Kansal}
\affiliation{Physics of Fluids Group, Faculty of Science and Technology, University of Twente, P.O. Box 217, 7500AE Enschede, Netherlands}
\author{Miguel A. Herrada}
\affiliation{Depto. de Mec{\'a}nica de Fluidos e Ingenier{\'i}a Aeroespacial,
Universidad de Sevilla, E-41092 Sevilla, Spain}
\author{Jens Eggers}
\affiliation{School of Mathematics,
University of Bristol, Fry Building, Woodland Road, Bristol BS8 1UG, United Kingdom}
\author{Jacco H. Snoeijer}
\affiliation{Physics of Fluids Group, Faculty of Science and Technology, University of Twente, P.O. Box 217, 7500AE Enschede, Netherlands}

\date{\today}

\begin{abstract}
A slender thread of elastic hydrogel is susceptible to a surface instability that is reminiscent of the classical Rayleigh-Plateau instability of liquid jets. 
The final, highly nonlinear states that are observed in experiments arise from a competition between capillarity and large elastic deformations. Combining a slender analysis and fully three-dimensional numerical simulations, we present the phase map of all possible morphologies for an unstable neo-Hookean cylinder subjected to capillary forces. Interestingly, for softer cylinders we find the coexistence of two distinct configurations, namely, cylinders-on-a-string and beads-on-a-string. It is shown that for a given set of parameters, the final pattern is selected via a dynamical evolution. To capture this, we compute the dispersion relation and determine the characteristic wavelength of the dynamically selected profiles. The validity of the ``slender" results is confirmed via simulations and these results are consistent with experiments on elastic and viscoelastic threads.
\end{abstract}

\maketitle

\section{Introduction}

The breakup of a cylindrical liquid jet into small droplets is a paradigmatic example of capillary action~\cite{E97,EV08}. The associated Rayleigh-Plateau instability that triggers the break up process is a generic mechanism for droplet formation and is extensively used in applications involving sprays and printing~\cite{B02}. Remarkably, it was shown recently that \emph{solid} cylinders can also exhibit a Rayleigh-Plateau instability, albeit only below a critical value of the elastic modulus~\cite{mora10}. Spontaneous undulations appear on slender strands of soft agar gel, and lead to a periodic structure with large deformations.  The threshold of instability was computed to be when the shear modulus $\mu$ falls below $\frac{1}{6}\gamma/h_0$, where $\gamma$ and $h_0$ are the surface tension and cylinder radius, respectively, in close agreement with experiment~\cite{mora10}. This provides an elegant demonstration that elastic solids, when sufficiently soft, are shaped by capillarity in a way similar to liquid interfaces. Indeed, the effect of surface tension has turned out to be essential for adhesion and wetting of soft solids~\citep{Style13_1, Chakrabarti18_1, Snoeijer20}, flattening of sharp features on soft surfaces~\cite{paretkar14}, and interaction and self-assembly on soft, deformable substrates~\citep{Maha2015, Chakrabarti2014, Pandey_2018, Karpitschka2016}.

\begin{figure*}[t]
    \centering
    \includegraphics[width=1.0\textwidth]{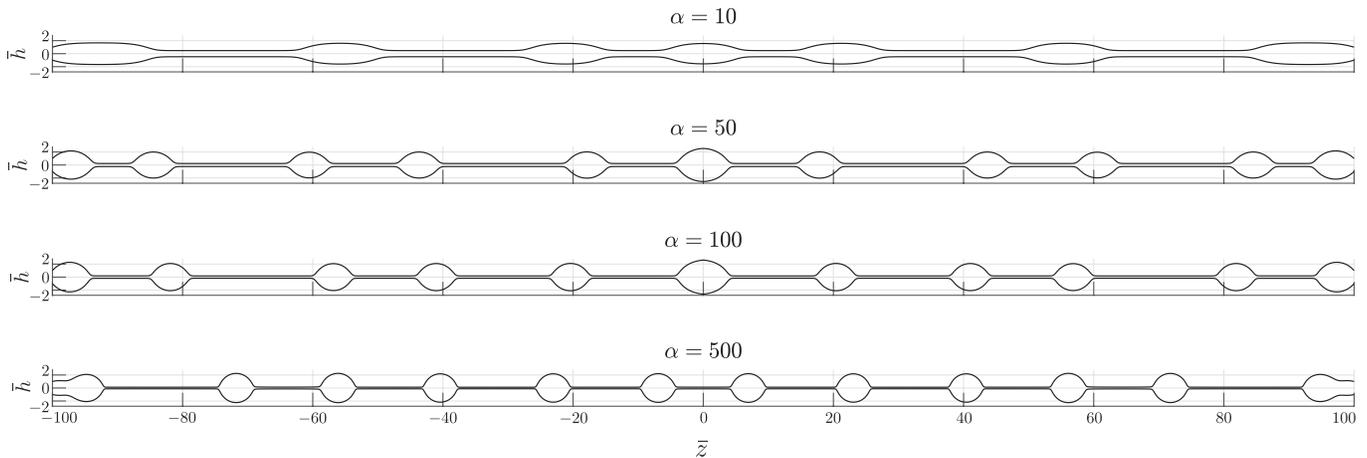}
    \caption{Typical stationary states resulting from the elastic Rayleigh-Plateau instability. These shapes are obtained from our dynamical numerical simulations of elastic neo-Hookean cylinders subjected to capillarity (without invoking a slender assumption). The initial condition is a long cylinder of uniform radius, but small numerical errors become amplified by a linear instability to produce a non-uniform state. From top to bottom, the panels correspond to cylinders of deceasing stiffness, characterised by $\alpha=\gamma/(\mu h_0)$.  The thread initially has a dimensionless radius $\bar h=h/h_0=1$, and extends over a finite domain from $\bar z=z/h_0 = -100$ to $100$. The domain was taken very long such that the finite domain-size has a limited effect on the wavelength selected in the final pattern. Note that the simulations are symmetric around $\bar z=0$.  
              \label{fig:numerics}}
\end{figure*}

Contrary to their liquid counterparts, however, elastic cylinders do not
pinch off to form droplets. Instead, they evolve towards static shapes that are characterized by large elastic deformations, where surface tension balances the nonlinear elastic stress. As an illustration of typical stationary 
profiles, we already provide some of our numerical results in
Fig.~\ref{fig:numerics}, which will be described in detail in the course
of the manuscript. The initial condition is a long cylinder of uniform radius, but small numerical errors become amplified by a linear instability to produce a non-uniform state. The various profiles in Fig.~\ref{fig:numerics} correspond to different stiffnesses, characterised by the dimensionless number $\alpha=\gamma/(\mu h_0)$. In all cases, the final state of the thread consists of thick bulges that are connected by thin threads. We wish to describe these profiles in detail, in particular how their size and spacing depend on the stiffness.

The bifurcation scenario of the instability and the subsequent nonlinear states have been analyzed using  neo-Hookean solids augmented with a surface tension~\cite{Ciarletta15,biggins_2017,audoly_arxiv2020}. The bifurcations are intricate. For example, the critical shear modulus can be adjusted by pre-stretching the thread~\cite{Ciarletta15}. At moderate stiffness (upper panel in Fig.~\ref{fig:numerics}), the resulting nonlinear states are reasonably well described by thin cylinders that are connected to thicker cylinders. Indeed, it was found that elastocapillary cylinders of different radii can coexist below a critical stiffness~\cite{biggins_2017}, in a process that is analogous to e.g. the coexistence of a liquid and vapour phases. The cylinders of different radii are connected by a transition region whose properties have been characterised~\cite{biggins_2017,audoly_arxiv2020}.

Interestingly, similar morphologies are encountered during the breakup of
\emph{viscoelastic liquids}, for example solutions of long,
flexible polymers. Typical configurations are jets of fluid emanating
from a nozzle \cite{CEFLM06}, drops held between two plates sufficiently
far apart for the liquid bridge to be unstable \cite{CEFLM06}, or drops
dripping from a pipette \cite{WABE05,DHEB20}. During breakup, long
threads of liquid are observed, which become thinner over time~\cite{AM01},
while most of the liquid is collected into almost spherical drops,
to form the `beads-on-a-string' (BOAS) configuration: large drops
connected by thin uniform threads~\cite{CEFLM06}. The stretching of long
flexible polymers in the thin liquid threads inhibits a rapid drop
pinch-off. When modelling the process using an Oldroyd-B liquid, the
thickness of the thread decays exponentially in time with a timescale
given by the relaxation time of the
polymer~\citep{BER90, CEFLM06, DVB18,Yacine01}. From the thinning dynamics,
one can therefore infer the extensional rheology of the viscoelastic
liquid~\cite{MS_rev,eggers_herrada_snoeijer_2020}.

While the nonlinear states of viscoelastic liquids exhibit a BOAS
morphology, the experimental shapes on the agar gels more closely resemble `cylinders-on-a-string' (COAS), in line with analysis~\cite{biggins_2017}. One might be inclined to think that this difference is due to the different types of material; one being dilute polymer suspensions while the solid hydrogels consist of a fully crosslinked polymer network that is swollen by water. However, the images toward the bottom of Fig.~\ref{fig:numerics}, corresponding to very small stiffness, one also observes BOAS in our purely elastic numerical simulations. Indeed, in the limit of large relaxation times, the elastic stress of the Oldroyd-B fluid is described exactly by a neo-Hookean solid \cite{Larson88,ES_elastic}. In the context of axisymmetric jets it was shown in a lubrication description that the transient shapes during pinch-off in the Oldroyd-B fluid are described by slender neo-Hookean solids~\cite{EYa84,CEFLM06,TLEAD18,ZD18}, and this scenario was recently confirmed in a fully three-dimensional analysis~\cite{eggers_herrada_snoeijer_2020}: a purely elastic BOAS solution was found for neo-Hookean solids, which is identical to the similarity solution for the transient viscoelastic thinning. The connection between elasticity and viscoelasticity was also explored recently~\cite{ZD18}, again using the idea that elastic cylinders of different radii can coexist~\cite{biggins_2017}. 

In this paper, we address the question of what selects the final pattern in the elastic Rayleigh-Plateau instability, i.e. what determines whether beads or cylinders are formed. Combining numerical simulations and a slender analysis, we  reveal the complete set of nonlinear states of elastic Rayleigh-Plateau instability. The main result is that for a given set of parameters, COAS and BOAS morphologies can both exist. However, the final solution is selected dynamically, and is determined from the characteristic wavelength during the onset of the instability. 

The manuscript is organised as follows. In Section~\ref{sec:slender} we start by exploring the nonlinear final states by a slender theory based on energy minimisation. We compute the map of all possible solutions in Section~\ref{sec:morphologies}, and show that both COAS and BOAS solutions can coexist for a given value of the stiffness. The remaining question is thus to find the mechanism that selects the morphology. In Section~\ref{sec:dynamics} we demonstrate that this selection is a dynamical process: to that end we compute the dispersion relation of the elastic Rayleigh-Plateau instability, identifying the growth rates and the dominant wavenumber. It is shown that the fastest growing wavelength accurately predicts the selected nonlinear state, as found in numerical simulations. Finally, in Section~\ref{sec:conservation} we lift the slender assumption and generalise various results through conservation laws associated to translational invariance with respect to space and material coordinates. The paper closes with a Discussion in Section~\ref{sec:discussion}. 

\section{Slender formulation}\label{sec:slender}

\subsection{Free energy functional}

We start by developing an analytical model to understand the capillary driven deformation of an elastic cylinder. Let us consider an infinitely long elastic cylinder of initial radius $h_0$. The analysis bears a resemblance with~\cite{biggins_2017,ZD18,audoly_arxiv2020}, but with some differences that enable the appearance of both `beads' and `cylinders' -- this will be highlighted in particular in Sec.~\ref{sec:morphologies}. 
In line with experiment, the spatial modulations triggered by the capillary instability are assumed to be axisymmetric. Hence, the shape of the deformed cylinder is described by the jet radius $h(z)$, where $z$ indicates the position along the thread. 

The associated capillary energy follows from the surface area, and reads

\begin{equation}
\mathcal{F}_\gamma = 2\pi \int \mathrm{d}z  \,\gamma\, h \sqrt{1+  h'^{2}},
\label{}
\end{equation}
where $h'=dh/dz$. 
We consider the solid surface tension $\gamma$ to be constant, but note that results for variable surface tensions are easily generalised using the Shuttleworth relation~\cite{Shut50, AS16, Pandey20}. For liquid jets an instability appears since long wave modulations lead to a reduction of $\mathcal F_\gamma$. For solids, however, the modulations induce elastic deformations that come with an elastic energy

\begin{equation}
\mathcal F_e = 2\pi \int\int \mathrm{d}R\,\,\mathrm{d}Z\,\,  R\, W(\mathbf F).
\end{equation}
In this expression $W$ is the elastic energy density (per unit reference volume), integrated over the radial $R$ and axial $Z$ coordinates of the reference state, i.e. prior to deformation. The elastic energy depends on the deformation gradient tensor $\mathbf F=\partial\mathbf{x}/\partial\mathbf{X}$, which accounts for the mapping from the reference ($\mathbf{X}$) to the current ($\mathbf{x}$) configuration. For an axisymmetric deformation field, the mapping is given by $r=r(R, Z)$ and $z=z(R, Z)$. Whether or not the Rayleigh-Plateau instability appears depends on the relative strength of the reduction of the capillary energy as compared to the increased elastic energy. Below we will consider a neo-Hookean solid~\cite{ogden_book}, for which $W={\mu\over2}(\mathrm{Tr}(\mathbf{F\cdot F}^T)-3)$. Note that the elastic energy is expressed in reference (Lagrangian) coordinates, whereas the surface energy is naturally expressed in current (Eulerian) coordinates. To be able to incorporate these into a total free energy, we rewrite $\mathcal F_e$ in Eulerian coordinates. Since we consider the cylinder to be incompressible, the volume elements in both of the coordinate systems are same, so the energy density $W$ is the same in when expressed on the domains of the current and reference coordinates.


In what follows we simplify the elastic energy by making use of the ``slender" approximation, commonly used for (viscoelastic) liquid threads, where spatial modulations of $h(z)$ along the jet are slow~\citep{CEFLM06}. Though approximate, it has the major advantage that it enables a detailed analytical exploration and classification of possible nonlinear states. We will find that the slender assumption is valid nearly everywhere along the jet, and we will confirm later that the ``slender" results are indeed fully representative of the true elastocapillary problem. 

In the slender approximation the axial stretch $\lambda$, defined as $\mathrm{d}z=\lambda\,\mathrm{d}Z$, will to leading order be homogeneous across the cross-section of the cylinder; for corrections including gradients of stretch we refer to\cite{audoly_arxiv2020}. Therefore, it can be written as a function of the axial coordinate alone, i.e. $\lambda(z)$, where we prefer to use the current axial position $z$ as the coordinate. For an incompressible material, the volume is preserved which means that an axial stretch must be counteracted by a reduction of the cross-sectional area. By consequence, the current radial position of a point of reference coordinate $R$ becomes $r=R/\sqrt{\lambda(z)}$, which fixes the radial stretching in terms of the axial $\lambda(z)$. In particular, for the radius of the interface this implies $h=h_0/\sqrt{\lambda}$. To be consistent with the slender approximation, we evaluate $\mathbf{F}$ to leading order and find that the principal values are given by $\lambda=h_0^2/h^2$ (axial direction) and $\lambda^{-1/2},\,\lambda^{-1/2}$ (radial and azimuthal directions). This provides a major simplification, also used in\cite{biggins_2017,audoly_arxiv2020}:  it allows expressing the elastic energy directly in terms of $h(z)$. Integrating $\mathcal F_e$ over the cross-section, we obtain the total elastocapillary energy:

\begin{equation}\begin{split}
\label{eq:energy}
\mathcal{F}[h] &= \pi \int \mathrm{d}z  \left(2 \gamma h \sqrt{1+  h'^{2}} + h^2 W(h) - p \,h^2\right) \\
&= \pi \int \mathrm{d}z  \left(2 \gamma h \sqrt{1+  h'^{2}} + w(h) - p\,h^2\right).
\end{split}\end{equation}
Here we introduced $w(h)\equiv h^2 W(h) $ as the elastic energy per unit length of the cylinder. For a neo-Hookean solid, the explicit form reads

\begin{equation}\begin{split}\label{eq:wh}
w(h) = h^2 W(h) &= \frac{h^2}{2}\mu \left( \lambda^2+{2\over\lambda}-3\right)\\
 &= {1\over 2}\mu \left( {h_0^4\over h^2}+2{h^4\over h_0^2}-3 h^2 \right).
\end{split}\end{equation}
In (\ref{eq:energy}) we furthermore introduced $p$ as a Lagrange multiplier to impose the volume constraint, accounting for the incompressibility of the material.

\begin{figure*}[t]
    \centering
    \includegraphics[width=0.80\textwidth]{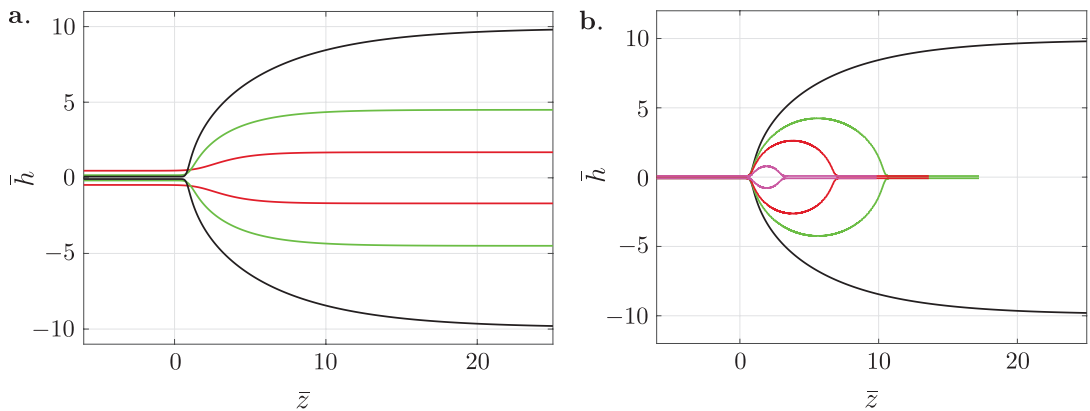}
    \caption{Different nonlinear morphologies of elastocapillary cylinders. (a) COAS shapes for increasing $\alpha$ ($10.08$ (Red), $102$ (Green), and $997$(Black)). The sizes of the cylinders are represented by $\times$ symbols in Fig.~\ref{fig:phase_plot}. (b) For $\alpha=997$, a gradual increase in $\bar{p}$ results in solutions that reach a constant curvature at the thick part giving BOAS configurations. The thread size is identical for these solutions following the scaling of \eqref{h_th}. The sizes of the beads are represented by open diamond symbols in Fig.~\ref{fig:phase_plot}.
    }
    \label{fig:profiles}
\end{figure*}

\subsection{Free energy minimisation}\label{fem}

We have now reduced the elasto-capillary free energy to \eqref{eq:energy}, which is a functional of $h(z)$. One notices that the integrand only involves $h$ and $h'$, so that the minimisation can be performed using the Euler-Lagrange equation, as e.g. in classical mechanics. Hence, we can introduce the effective ``Lagrangian" (omitting a factor $\pi$),

\begin{equation}
\mathcal L(h,h') = 2 \gamma h \sqrt{1+  h'^{2}} + w(h) -  p\,h^2.
\end{equation}
The corresponding Euler-Lagrange equation gives an ODE for the shape of the cylinder,  

\begin{equation}
p = \gamma \kappa + \frac{1}{2 h} \frac{\partial w}{\partial h}, \quad \mathrm{with} \quad \kappa = \frac{1}{h\sqrt{1+ h'^{2}}} -  \frac{h''}{(1+h'^{2})^{3/2}}.
\label{EL}     
\end{equation} 
Here we recognize the Laplace pressure due to the interface curvature $\kappa$, which is now complemented by an elastic contribution. Given that $\mathcal L$ does not depend explicitly on $z$, we can use the Beltrami identity\footnote{In the Lagrangian formulation of classical mechanics, the role of $z$ is played by time. In case the Lagrangian does not depend explicitly on time, the Beltrami identity implies that the Hamiltonian is constant in time and expresses conservation of energy.} to find an exact integral of \eqref{EL},

\begin{equation}
T =\Lag - h'\pd{\Lag}{h'}= \frac{2 \gamma  h} {\sqrt{1+h'^{2}}} + w(h) -  p\,h^2. 
\label{T} 
\end{equation}
The quantity $T$ represents the tension in the thread, the integral of elastic and capillary stress over the cross-section -- at equilibrium, this tension is indeed constant along $z$ to ensure a global force balance. In the energetic description, the ``conservation of tension" is thus a consequence of the invariance of the free energy with respect to a translation in $z$. This translational invariance is of course not restricted to the slender assumption; in Sec.~\ref{sec:conservation} we will find that a conservation law for both $T$ and $p$ can be derived beyond the slender assumption. 


In what follows, we work with dimensionless variables, $\bar{h}=h/h_0$, $\bar{z}=z/h_0$. The equilibrium equation thus takes the form

\begin{equation}
\bar{p}=\frac{1}{\bar{h} \sqrt{1+ \bar{h}'^{2}}} -  \frac{\bar{h}''}{(1+\bar{h}'^{2})^{3/2}}  + \alpha^{-1}\left(-\frac{1}{2  \bar{h}^4} + 2 \bar{h}^2\right),
\label{nd_P} 
\end{equation} 
which was also derived in \cite{CEFLM06,EYa84}, along with the tension equation

\begin{equation}
\bar{T}=\frac{2 \bar{h}} {\sqrt{1+\bar{h}'^{\,2}}} + \alpha^{-1} \left(\frac{1}{2 \bar{h}^2} + \bar{h}^4\right) - \bar p\,\bar{h}^2.
 \label{nd_T}
\end{equation}
Here $\bar{T}=T/\gamma h_0$, $\bar p=( p h_0/\gamma)+3/(2\alpha)$, and we introduced 

\begin{equation}
\alpha = \frac{\gamma}{\mu h_0}.
\end{equation}
The parameter $\alpha$ is the elastocapillary number, measuring the relative strength of surface tension to elasticity. It will be the central parameter describing the properties of the elastic thread.

\section{Morphologies}\label{sec:morphologies}

\begin{figure*}[t]
    \centering
    \includegraphics[width=.65\textwidth]{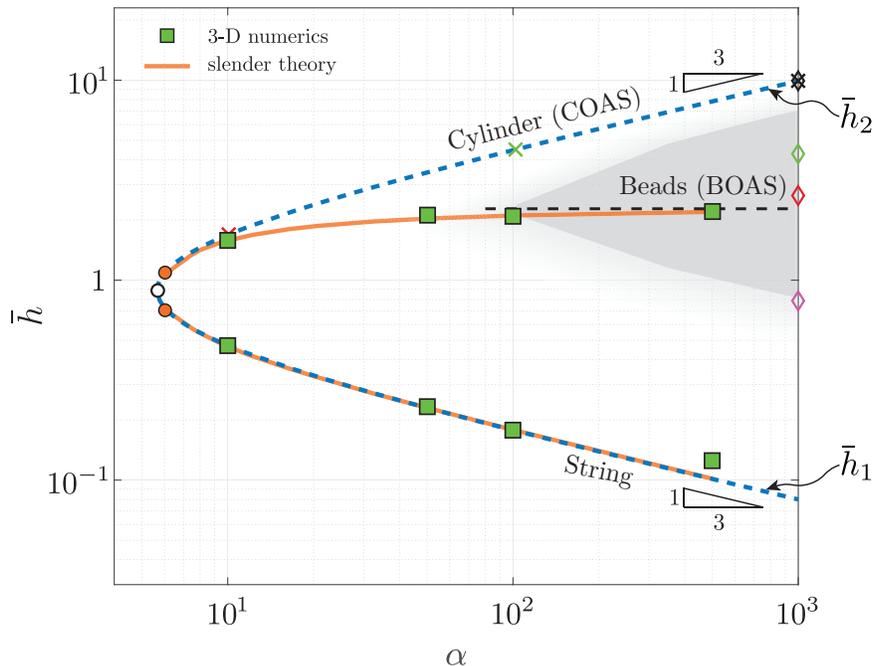}
    \caption{Phase map of the elastic Rayleigh-Plateau instability. Size of the bead, cylinder and string/thread are plotted as a function of the elastocapillary number. The blue dashed lines represent string ($\bar{h}_1$) and cylinder ($\bar{h}_2$) for COAS configuration. For large $\alpha$ these solutions obey the asymptotic results of \eqref{la_asmp}. The open circle marks the critical point given by $\alpha_c=\sqrt{32}$ and $\bar{h}_c=2^{-1/6}$. Beads (of sphericity greater than $60\%$) lie within the shaded (gray) region. The orange solid lines represent string size ($\bar{h}_1$) and maximum radius ($\bar{R}$) found from the slender theory for dynamically selected solutions conforming to Rayleigh-Plateau instability, for Ohnesorge number $\mathrm{Oh}=1$ (cf. section~\ref{sec:dynamics}). The orange circles represent onset of the dynamic instability ($\alpha=6$). For large $\alpha$, the bead size reaches an asymptotic limit given by eq.~\eqref{bead_asymp} and is plotted as the black, dashed line. The green squares mark the dynamically selected solutions found through 3-D numerics. The $\times$ symbols on blue dashed line ($\bar{h}_2$) represent the COAS cylinders from Fig.~\ref{fig:profiles}(a), while the open diamond symbols along the vertical line at $\alpha=997$ represent the sizes of the beads in Fig.~\ref{fig:profiles}(b), based on their respective colors.}
           \label{fig:phase_plot}
\end{figure*}

For a given set of $\bar{T}$ and $\bar p$ we can formally integrate \eqref{nd_T} by separation of variables, to find the deformed shape for any value of $\alpha$. In what follows, however, the integration is done numerically. Typical profiles are shown in Fig.~\ref{fig:profiles} confirming that indeed both the BOAS and COAS morphologies emerge within the present formulation, upon varying $\bar{T}$ and $\bar p$. Below we analytically derive the key features of the two configurations, \textit{e.g.} the sizes of the string, cylinder and bead. These results are partially available in the literature for either BOAS or COAS, but here we show how they emerge from a \emph{single} framework. This is important, as it reveals how these different morphologies are connected in the space of possible solutions. The combined results for BOAS and COAS will be summarised in a regime map (Fig.~\ref{fig:phase_plot}), offering a complete overview of the possible nonlinear states.

\subsection{Cylinders-on-a-string}

The experiments on Agar gel form nearly homogeneous cylinders with two different radii. We will denote the radius of thicker part of the cylinder by $\bar h_2$ which is connected to the thinner part of radius $\bar h_1$. Assuming uniform radii away from the transition region, the derivatives $\bar{h}'$ and $\bar{h}''$ vanish, and the expressions for $\bar p$ and $\bar{T}$ become algebraic functions of either $\bar h_1$ or $\bar h_2$. The selected pairs $(\bar h_1,\bar h_2)$ can be determined by noting that  both $\bar p$ and $\bar{T}$ are constant along the entire thread; $\bar p$ is a Lagrange multiplier for the total volume while $\bar T$ is an integration constant that represents the tension inside the thread. This means that we can evaluate (\ref{nd_P},\ref{nd_T}) at both $\bar h_1$ and $\bar h_2$, and equate the resulting values of $\bar p$ and $\bar T$, which gives two equalities:

\begin{subequations}
\begin{equation}
\frac{1}{\bar{h}_1} +{1\over\alpha}\left( 2\bar{h}_1^2-\frac{1}{2\bar{h}_1^4} \right)= \frac{1}{\bar{h}_2} + {1\over\alpha}\left(2 \bar{h}_2^2-\frac{1}{2\bar{h}_2^4}\right),
\label{eq_P}
\end{equation}
\begin{equation}
\bar{h}_1 +{1\over\alpha}\left( \frac{1}{\bar{h}_1^2} -\bar{h}_1^4\right) = \bar{h}_2 + {1\over\alpha}\left(\frac{1}{\bar{h}_2^2} - \bar{h}_2^4\right).
\label{eq_T}
\end{equation}
\label{eq_PT}
\end{subequations}
In Sect.~\ref{sec:conservation} below we will show that \eqref{eq_PT}
is in fact exact, without relying on lubrication ideas, in the limit that
the cylindrical parts are perfectly uniform.

For a given value of $\alpha$, these provide two equations that uniquely determine the COAS-pair ($\bar h_1,\bar h_2)$. Subsequently, we can evaluate the corresponding values of $\bar{p}$ and $\bar{T}$, and compute the entire profile of the thread. Before discussing the results, we note that (\ref{eq_P},\ref{eq_T}) are the exact same conditions as derived by Xuan and Biggins~\cite{biggins_2017} and Zhou and Doi~\cite{ZD18}, who used a Lagrangian description and identified the conditions for ``cylinder coexistence" based on an analogy with phase separation. 

The results for $\bar h_1$ and $\bar h_2$ are plotted in Fig.~\ref{fig:phase_plot} as the blue dashed lines, where the top branch represents $\bar{h}_2$ and the bottom branch gives $\bar{h}_1$. Clearly, nontrivial solutions only appear above a critical $\alpha_c$, marked by the open circle in Fig.~\ref{fig:phase_plot} . This value of $\alpha$ is found by noting that two branches merge at the critical point, which appears at the cylinder radius of $2^{-1/6}$ at $\alpha_c=\sqrt{32}$, as previously pointed out by Taffetani and Ciarletta~\cite{Ciarletta15}, Xuan and Biggins~\cite{biggins_2017}, and Zhou and Doi~\cite{ZD18}. With increasing $\alpha$ the thread ($\bar h_1$) becomes asymptotically thin, while the cylinder ($\bar h_2$) progressively bulges out. In the limit of large $\alpha$, corresponding to very soft cylinders, one can derive the following asymptotic expressions~\cite{ZD18}

\begin{equation}
\bar h_ 1 = (2\alpha)^{-1/3}, \quad \quad \bar h_2 = \alpha^{1/3}.
\label{la_asmp}
\end{equation}

An interesting feature already noted in~\cite{Ciarletta15,biggins_2017} is that the critical point $\alpha_c$ lies below the onset of instability of an \emph{unstretched} elastic cylinder, which is at $\alpha=6$ as found from linear stability analysis~\cite{mora10}. The values for $(\bar h_1,\bar h_2)$ at $\alpha=6$ are indicated by the orange circles in Fig.~\ref{fig:phase_plot}. This means that for an unstreched cylinder, the transition to COAS is discontinuous, reaching a finite difference amplitude already at the transition. A continuous transition is recovered when stretching the cylinder such that it matches the radius of the critical point, $\bar h= 2^{-1/6}$.

A great asset of the slender approximation is that it allows for computing the full shape of the soluion by direct integration of \eqref{nd_T}. For a chosen $\alpha$, the string thickness ($\bar{h}_1$) is found by solving eqns.~\eqref{eq_P} and \eqref{eq_T}, and the values of $\bar T$ and $\bar p$ are determined subsequently. The profiles follow from numerical integration of the resulting first order ordinary differential equation \eqref{nd_T} using NDSolve in Mathematica. The results are shown in Fig.~\ref{fig:profiles}a. For large $\alpha$ we clearly observe the COAS morphology. Upon decreasing $\alpha$ the transition region between the two cylinders becomes smoother, with an increasing width near the critical point~\cite{biggins_2017}.
 
\subsection{Beads-on-a-string}

The possibility of spherical beads, as commonly observed for viscoelastic liquids \cite{CEFLM06, BAHPMB10}, is not captured in Xuan and Biggins~\cite{biggins_2017}. The reason for this is that the axial curvature was not incorporated into the description, due to an approximation of the surface energy. Conversely, the usual description for BOAS in viscoelastic liquids \cite{EF_book,CEFLM06} does not capture the COAS solutions described above, as the elastic terms associated to radial stretching were dropped in the asymptotic analysis (specifically, the term $\sim h^4$ in $w(h)$). Here we show how the beads occur within the present formalism, which thus contains both BOAS and COAS. As such, we can predict when these morphologies are expected to occur in the elastic Rayleigh-Plateau instability.

First we show when spherical beads of size $\bar{R}$, with constant curvature $\bar{\kappa}=2/\bar{R}$, can appear as a solution. The pressure equation (\ref{nd_P}) indeed allows for constant curvature solutions, provided that the elastic terms proportional to $\alpha^{-1}$ are subdominant. This implies that $1/\bar{R} \gg \alpha^{-1}\bar{R}^2$, which provides an upper bound $\bar{R} \ll \alpha^{1/3}$. With this, $\bar p \simeq 2/\bar{R}$, and we can once again make use of the fact that $\bar p$ and $\bar T$ are constant everywhere along the jet. Evaluating \eqref{nd_P} inside the cylindrical thread of radius $\bar h_{\rm th}$, we obtain

\begin{equation}
\frac{2}{\bar{R}} = \frac{1}{\bar{h}_{\rm th}} - \frac{1}{2 \alpha} \frac{1}{\bar{h}_{\rm th}^4} + \frac{2}{\alpha} \bar{h}_{\rm th}^2,  
\end{equation}
If in addition $\bar R$ is large compared to the thread radius $\bar{h}_{\rm th}$, the leading order balance reads

\begin{equation}
0 = \frac{1}{\bar{h}_{\rm th}} - \frac{1}{2 \alpha} \frac{1}{\bar{h}_{\rm th}^4} \quad \Rightarrow \quad \bar{h}_{\rm th} = (2 \alpha) ^{-1/3}.
\label{h_th}
\end{equation}
The analysis is based on the hierarchy of scales $\bar h_{\rm th}\sim \alpha^{-1/3} \ll \bar R \ll \alpha^{1/3}$. Hence, the BOAS solutions can only appear at large $\alpha$. For completeness, we can also compute the tension associate to this solution, which gives $\bar{T}= 3/(2 \alpha)^{1/3}$, in agreement with \cite{CEFLM06, EF_book}.
 
Importantly, and perhaps somewhat unexpectedly, the thread radius $\bar h_{\rm th}$ for the BOAS solution, equation (\ref{h_th}), converges to the exact same thickness as the large $\alpha$ asymptotics of $\bar h_1$ for the COAS, reported in (\ref{la_asmp}). This was also noticed in~\cite{ZD18}. Hence, the thread thickness is universal in the limit of very soft cylinders. It is therefore of interest to investigate what the various solutions look like, by integrating (\ref{nd_T}) under the conditions that a thin thread appears. This is achieved by considering small but finite values of $\bar p$; varying $\bar p$ effectively controls the volume of the solutions. The result is given in  Fig.~\ref{fig:profiles}b, for the case where $\alpha=997$. One observes that the string can be connected to beads of arbitrary size, until the bead size reaches the upper bound $\bar{R} \sim \alpha^{1/3}$. Then, the size of the beads ``saturates" and gives rise to a flat cylinder, shown by the black profile in Fig.~\ref{fig:profiles}b. The reason for this saturation is that the elastic energy associated to the radial expansion, required to form large beads, becomes comparable to the capillary energy. 

\subsection{Regime map}

These findings are summarised in the regime map given in Fig.~\ref{fig:phase_plot}. The COAS solutions are indicated by the blue dashed lines, indicating the thickness of the cylinders as a function of $\alpha$. At large $\alpha$, the thin thread can equally well be connected to spherical beads as long as the bead radius $\bar{R}$ remains below the saturation given by the blue line. The region where BOAS can be observed is indicated by the grey zone, which for large $\alpha$ spans a very large part of the solution space. The grey zone is not sharply defined: it is drawn by evaluating the ``sphericity" of the beads in the numerical profile (defined somewhat arbitrarily by demanding $\kappa h_{\rm max}/2$ to be larger than $\sim 60\%$). We thus conclude that for moderate values of the softness, $\alpha \sim \mathcal O(10)$, only cylinder-like solutions can be observed. This is consistent with experiments~\cite{mora10} and previous analysis~\cite{biggins_2017}. By contrast, both BOAS and COAS can appear at larger values of $\alpha$. 

The remaining question is, for a given value of $\alpha$, which of these solutions will be selected in an experiment, or in the numerical simulations presented above. For each of the numerical solutions in Fig.~\ref{fig:numerics}, to be discussed in detail below, we determined the thread radius and the maximum height along the cylinder (averaged over the various beads). The resulting values are added to the regime map of Fig.~\ref{fig:phase_plot} as green squares. The threads in the numerical simulations very closely follow the prediction for $\bar h_1$, indicated by ``string", obtained from the cylinder analysis. However, the size of the bulges are typically much below the value of $\bar h_2$, indicated by ``cylinder": the sizes tend towards the grey region corresponding to BOAS. Below we explain the mechanism that selects the bead size and derive the orange lines in Fig.~\ref{fig:phase_plot}.

\section{Dynamical selection of bead size}\label{sec:dynamics}

In the Rayleigh-Plateau instability of a liquid, one can obtain a reasonable estimate of the size of the large drops from the dispersion relation of the linear stability analysis. The fastest growing wavelength dictates the periodicity of the initial pattern, which combined with volume conservation gives an estimate of the drops. We propose the same scenario for the selection of the bead size in the elastic Rayleigh-Plateau instability.

Below we spell out the dynamical equations, to trace the temporal
evolution from an initially homogeneous cylinder, subject only to small
perturbations such as those resulting from numerical rounding error, 
to the final nonlinear states. We discuss numerical solutions to these
equations, and also determine the dispersion relation by computing the
growth rate of small sinusoidal perturbations 
(with and without the slender assumption). Subsequently, the fastest growing wavelength is used to estimate the bead size, and show how this provides the selection of the nonlinear state in the regime map of Fig.~\ref{fig:phase_plot}.

\subsection{Numerical solution}\label{sec:num}

\subsubsection{Dynamical constitutive equation}

To determine the cylinder's evolution, we need to extend the energetic formulation to include the dynamics of the instability. We here formulate the fully three-dimensional problem and present numerical solutions. In the Eulerian formulation for an incompressible medium, this is given by the mass and momentum equations 

\begin{eqnarray}
\label{equ1}
 \nabla \cdot{\bf v}& =& 0,\\
\label{equ2}
 \rho\left(\frac{\partial {\bf v}}{\partial t} +
\left({\bf v}\cdot \nabla \right){\bf v} \right)
&=& \nabla \cdot\mathbf\sigma,
\end{eqnarray}
where $\rho$ is the density, ${\bf v}$ the velocity field and $\sigma$ is the Cauchy stress tensor. For a neo-Hookean solid the Cauchy stress reads $\mathbf\sigma=\mu \mathbf F\cdot \mathbf F^T-p\mathbf I$, where we remind $\mathbf F=\partial \mathbf x/\partial \mathbf X$ is the deformation gradient due to the mapping from reference ($\mathbf X$) to current ($\mathbf x$) configuration. Since we are now dealing with the dynamical evolution, the mapping $\bf{x} (\bf{X},t)$ is time-dependent. To relate the mapping, needed for the neo-Hookean stress, and the velocity $\mathbf v$, we use~\cite{Kamrin2012},

\begin{equation}
\frac{\partial {\bf X}}{\partial t} + {\bf v}\cdot\nabla{\bf X} = 0,
\label{reference}
\end{equation}
where $\bf{X}(\bf{x},t)$ is the inverse of the mapping. 

Importantly, the dynamics (\ref{equ2}) for a neo-Hookean solid is purely conservative: to reach a thread at a stationary state, one needs to include mechanism that dissipates the elasto-capillary energy that is liberated after the instability sets in. In experiment, the dissipative mechanism is material specific, and could e.g. be rather different when comparing elastomeric and hydrogel cylinders. Here we choose to simply add a (Newtonian) viscous contribution to the stress, so that

\begin{equation} \label{eq:numstress}
\mathbf{\sigma}=-p {\bf I}+ \eta\left(\nabla {\bf v}+\nabla{\bf v}^T\right)+ \mu\left({\bf F}\cdot{\bf F}^{T}-{\bf I}\right),
\end{equation}
where $\eta$ is the viscosity. This choice is natural in relation to the BOAS formation observed for viscoelastic liquids. Namely, as shown in \cite{ES_elastic}, the constitutive relation (\ref{eq:numstress}) corresponds to the Oldroyd-B model for viscoelastic liquids, taken in the limit of infinite relaxation times. Once a steady-state is reached, the viscous term vanishes and (\ref{eq:numstress}) reduces to the neo-Hookean solid. 

Below we report the results in dimensionless form. All lengths are again expressed in terms of $h_0$, and the ratio of elasticity to capillarity expressed by $\alpha=\gamma/(\mu h_0)$. The appearance of inertia now gives rise to a capillary timescale $(\rho h_0^3/\gamma)^{1/2}$, which will be used to scale the growth-rates. The relative importance of viscous dissipation is then expressed via the dimensionless group, 

\begin{equation}
\mathrm{Oh} = \frac{\eta}{\sqrt{\rho \gamma h_0}},
\end{equation}
which is the Ohnesorge number. 

\subsubsection{Numerical method}

We numerically solve the fully three-dimensional problem, but make use of the fact that profiles are axisymmetric. 
We then write the problem in cylindrical coordinates $(R,Z)$ and $(r,z)$, respectively corresponding to reference and current configurations. 
The corresponding deformation tensor is
$\mathbf{F} = (\partial r/\partial R){\bf e}_r\otimes{\bf e}_r +
(\partial r/\partial Z){\bf e}_r\otimes{\bf e}_z + (\partial z/\partial R){\bf e}_z\otimes{\bf e}_r +(\partial z/\partial Z){\bf e}_z\otimes{\bf e}_z
+ ( r/ R){\bf e}_\theta\otimes{\bf e}_\theta$. 
To improve the stability of the method, it is convenient to replace the incompressibility equation (\ref{equ1})  by 

\begin{equation}
\det{\bf F} = \frac{r}{R}\left(\frac{\partial r}{\partial R}
\frac{\partial z}{\partial Z}-\frac{\partial r}{\partial Z}
\frac{\partial z}{\partial R}\right)= 1.
\label{incompr}
\end{equation}
However, the stress boundary conditions at the free surface are expressed in their Eulerian form

\begin{equation}
{\bf n}\cdot\mathbf{\sigma} =
-\gamma\kappa{\bf n},
\label{stress_bc}
\end{equation}
where $\kappa$ is again (twice) the mean curvature and $\mathbf n$ the surface normal. 
Denoting $h(z,t)$ as the thread profile, the kinematic boundary conditions becomes

\begin{equation}
\frac{\partial h}{\partial t} + v_z(h,z)\frac{\partial h}{\partial z} =
v_r(h,z).
\label{kin_bc}
\end{equation}
At  $z=0$ we impose symmetry conditions 
\begin{equation}
v_{z}=0,\quad \frac{\partial v_r}{\partial z}=\frac{\partial h}{\partial z}=0,\label{symmetry}
\end{equation}
while regularity at $r=0$ requires

\begin{equation}
 v_{r}= \frac{\partial v_r}{\partial r}=0.\label{axis}
\end{equation}
Owing to the symmetry, the problem is resolved only on the the domain from $z=0$ to $z=L$. 

We have performed two types of simulations. On the one hand, we simulate cylinders of large but finite length. This is done by introducing no-slip boundary conditions at $z= L$, 

\begin{equation}
v_{z}=v_{r}=0,\quad h=h_0. \label{1Dbasic}
\end{equation}
We are interested in observing the natural wavelength that is selected by the nonlinear evolution equations. To restrict effects due to a finite length of the cylinder, we therefore take a large domain, $L=100 h_0$, which admits a broad spectrum of wavelengths to develop. On the other hand, we performed a numerical linear stability analysis for an infinitely long cylinder. Both problems are solved numerically using a mapping technique that is a variation of \cite{Herrada2016} for interfacial flows and extended in \cite{eggers_herrada_snoeijer_2020} for viscoelastic fluids. Details are provided in Appendix~\ref{app:mapping}. 

\subsubsection{Numerical results}

Figure~\ref{fig:numerics} shows the stationary states achieved for decreasing stiffness, from $\alpha=10$ to $\alpha=500$. The Ohnesorge number was kept constant, at $\mathrm{Oh}=1$. Each of the simulations starts from a homogeneous cylinder with vanishing elastic stress, whose small
numerical perturbations are amplified by a linear instability. 
The instability is
triggered by a release of elasto-capillary energy, which initially is
converted into motion. Owing to the viscous term, this energy is
dissipated and leads to the steady patterns shown in  Fig.~\ref{fig:numerics}. 

Clearly, the thickness of the thin thread decreases when the material becomes softer, i.e. with increasing $\alpha$. The thread thickness measured from the simulations is plotted in the regime map of Fig.~\ref{fig:phase_plot} as green squares, and it very closely follows the predicted thickness of the slender theory, indicated by ``string". The size of the bulges are also reported as green squares. For $\alpha=10$, which is within the experimental range for Agar gels~\cite{mora10}, the bulge size is close to the ``cylinder" analysis by~\cite{biggins_2017}. However, for large $\alpha$ the bulges are much smaller than predicted in the cylinder analysis, and we approach the BOAS regime. 

The simulations in Fig.~\ref{fig:numerics} were performed on a large domain to allow for the natural selection of a wavelength of the final pattern. Though the restrictions by the finite size lead to a variable spacing between the drops, we clearly observe a decrease in characteristic wavelength as $\alpha$ increases. To further investigate this, we numerically determined the dispersion relation by a linear stability analysis of the dynamical system, by computing the growth rates $\omega$ for an infinitely long cylinder, with a superimposed
  sinusoidal perturbation of small amplitude. We refer to Appendix~\ref{app:mapping} for details. 

\begin{figure}[t]
    \centering
    \includegraphics[width=\linewidth]{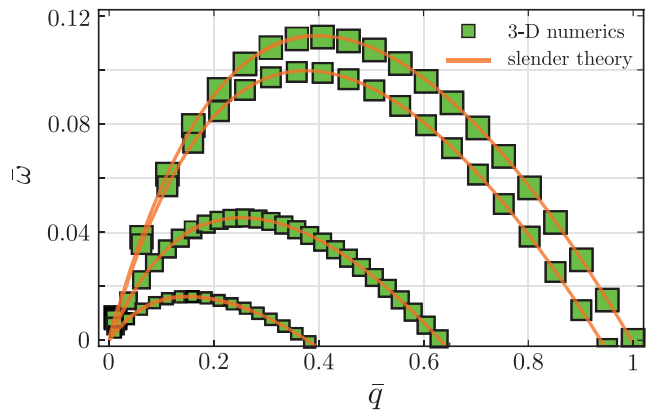}
    \caption{Dispersion relation, theory vs numerics: green squares represent results from 3-D simulations with increasing marker size denoting data for larger $\alpha$ (7, 10, 50, 1000). The Ohnesorge number is kept fixed at 1. The orange lines are the theoretical prediction given by \eqref{eq:dispersion}.}
    \label{fig:dispersion}
\end{figure}

The results for $\rm{Oh}=1$ are shown as the symbols in Fig.~\ref{fig:dispersion}, for varying $\alpha$. The results are plotted in dimensionless form, using $\bar \omega = \omega (\gamma/\rho h_0^3)^{-1/2}$ and $\bar q=qh_0$. Indeed, we see that the wavenumber with maximum $\bar \omega$ increases with $\alpha$, which is consistent with a decreasing wavelength of the pattern. These trends will be commented in detail below using analytical results from a slender dynamical analysis.

\subsection{Fastest growing wavelength in the slender approximation}

We now analyse the dynamical equations (\ref{equ1},\ref{equ2}) in the slender approximation. The approach is inspired by the slender formulation for viscoelastic fluids~\cite{EF_book} , but taken for infinite relaxation time to account for the neo-Hookean character. The slender (lubrication) approximation for thin threads requires two fields: the thread radius $h(z,t)$ and the axial component of velocity $v(z,t)\equiv v_z(z,t)$. In the slender approximation, these fields depend only on the axial coordinate $z$.   Mass conservation then gives 

\begin{equation}\label{eq:mass}
\frac{\partial h^2}{\partial t} + \left( h^2 v\right)'=0,
\end{equation}
which is complemented by the axial momentum conservation, 

\begin{equation}
\rho\left( \frac{\partial v}{\partial t} + v v'\right) 
= - \left( \frac{1}{2\pi h} \frac{\delta \mathcal F}{\delta h} \right)' + 3\eta \frac{\left( h^2 v'\right)'}{h^2}.
\end{equation}
The first term on the right can be viewed as the pressure gradient driving the flow, containing both capillarity and elasticity. Here we explicitly write it in a variational form, based on the functional derivative of $\mathcal F$. This offers a natural way to connect to the energy minimisation scheme that is employed in the first part of the paper, where we computed the stationary states. Working out the functional derivative, the momentum equation becomes

\begin{equation}\label{eq:momentum}
\rho\left( \frac{\partial v}{\partial t} + v v'\right) = - \left(\gamma \kappa + \frac{1}{2h}\frac{dw}{dh} \right)' + 3\eta \frac{\left( h^2 v'\right)'}{h^2}.
\end{equation}
When combined with (\ref{eq:wh}), the above equation is nearly of the same form as the lubrication equation in \cite{EF_book}, used for the Oldroyd-B fluid with infinite relaxation time. The only difference is that in (\ref{eq:wh}) we retained the term $\sim h^4$ in $w(h)$. This term is negligible in the BOAS regime, but not in general. Specifically, it turns out to be critical to explain the threshold of the elastic Rayleigh-Plateau instability.

We now proceed via a standard linear stability analysis. For this we introduce dimensionless variables, again scaling all lengths on $h_0$ and introducing the dimensionless wavenumber $\bar q = q h_0$. Similarly, we define the dimensionless growth rate $\bar \omega = \omega (\gamma/\rho h_0^3)^{-1/2}$ based on the (inverse) inertia-capillary time. We then consider small perturbations,

\begin{equation}
\bar h(\bar z,\bar t) = \left( 1+ \epsilon e^{\bar \omega \bar t} \cos \bar q \bar z\right), \quad 
\bar v(\bar z,\bar t) = \bar v_0 \epsilon e^{\bar \omega \bar t} \sin \bar q\bar z,
\end{equation}
retaining only terms linear in $\epsilon$. Mass conservation (\ref{eq:mass}) then gives the relation $\bar v_0 = -2\bar \omega/\bar q$, which is inserted in (\ref{eq:momentum}) to give the dispersion relation, 

\begin{equation}
\bar \omega^2 = \frac{1}{2}\left[ \bar q^2\left(1- 6\alpha^{-1}\right) - \bar q^4 \right] - 3 \mathrm{Oh} \, \bar \omega \bar q^2.
\label{eq:dispersion}
\end{equation}
We remark that for vanishing elasticity ($\alpha=\infty$), the dispersion relation of~\eqref{eq:dispersion} reduces to that of a slender, viscous jet~\cite{EV08}.

The solid lines in Fig.~\ref{fig:dispersion} show the dispersion relation (\ref{eq:dispersion}), for $\mathrm{Oh}=1$ and different values of the softness, $\alpha$. Clearly, the slender prediction is an excellent approximation over the entire range of $\alpha$. Hence, in the remainder we can work with a closed-form analytical expression for the fastest growing wavenumber $\bar q^*$, as given by (\ref{eq:star}).

\begin{figure*}[t]
    \centering
    \includegraphics[width=.8\linewidth]{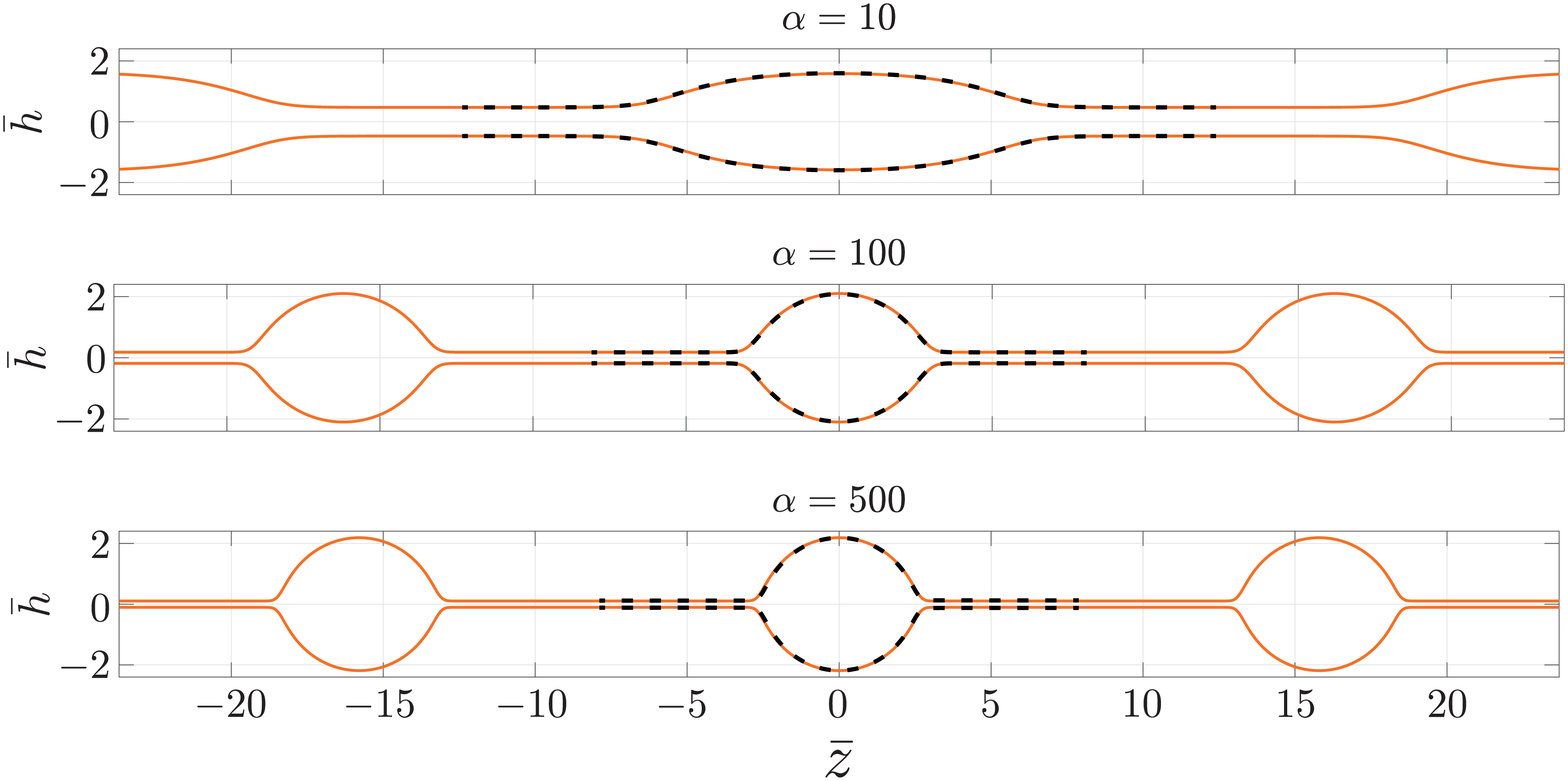}
    \caption{Comparison of dynamically selected profiles obtained by 3-D numerics (black dashed lines) and by slender theory (orange solid lines). These shapes correspond to the green squares of Fig.~\ref{fig:phase_plot}.}
    \label{fig:dyn_profiles}
\end{figure*}

Inspecting~\eqref{eq:dispersion} one verifies that the instability appears for $\alpha>6$, in agreement with the purely energetic analysis~\cite{mora10,Ciarletta15}. Unstable modes are found in the range $\bar q \in [0,\bar q_m]$, where $\bar q_m = \left( 1- 6 \alpha^{-1}\right)^{1/2}$, while the fastest growing mode, i.e. with the largest $\mathrm{Re}(\bar \omega)$, corresponds to 


\begin{equation}\label{eq:star}
\bar q^* = \frac{ \left( 1- 6 \alpha^{-1}\right)^{1/2}}{\left( 2+ 3 \sqrt{2} \mathrm{Oh} \right)^{1/2}}.
\end{equation}
As $\alpha \rightarrow 6^+$, we find that $q_m \rightarrow 0$. Hence, the onset of instability is governed by very long waves. On the other hand for $\alpha \rightarrow \infty $, elasticity ceases to be important and one recovers the usual Rayleigh-Plateau instability. A few comments are in order. First, we remark that, despite the slender assumption used here, the onset of the instability at $\alpha=6$ is in perfect agreement with the complete linear analysis \cite{mora10}. The reason for this is that the onset is at $\bar q=0$, for which the slender treatment of the elastic energy is asymptotically correct. Also in the limit $\alpha \rightarrow \infty$, the range of unstable wavenumbers again becomes exact. Namely, this limit is dictated by capillarity, which here is captured by the exact form of $\mathcal F_\gamma$. 


\subsection{Selection of bead size}

We assume that the periodicity of the final nonlinear state is dictated by the most unstable wavelength, $\lambda^* = 2\pi/q^*$. In addition, the original volume per wavelength, $\pi h_0^2\lambda^*$ must be preserved in the nonlinear state. For any value of $\alpha$, these two conditions select a unique pair of $\bar{p}$ and $\bar{T}$, so that a unique nonlinear solution can be determined from (\ref{nd_T}). 

To illustrate the selection mechanism, we first consider the limit of $\alpha \rightarrow \infty$, for which the volume inside the thin thread becomes negligible with respect to that inside the bead of radius $R$. In this extreme case, volume conservation thus takes the form

\begin{equation}
\pi h_0^2 \lambda^* = 2 \pi^2 h_0^3 \left( 2+ 3 \sqrt{2} \mathrm{Oh}\over1-6\alpha^{-1} \right)^{1/2}= \frac{4}{3}\pi  R^3,
\end{equation}
where we used $\lambda^*$ from the linear stability analysis. Hence, we find

\begin{equation}\label{eq:selection}
\frac{R}{h_0} = \bar R = \left[\frac{3}{2}\pi  \left( {2+ 3 \sqrt{2} \mathrm{Oh}\over1-6\alpha^{-1}} \right)^{1/2}\right]^{1/3}.
\end{equation}
This indeed provides a selection of a unique bead size, which approaches a constant value in the limit of large $\alpha$ (this limiting value is indicated for $\mathrm{Oh}=1$ in Fig.~\ref{fig:phase_plot}). The analysis is consistent as long as $\bar R \ll \alpha^{1/3}$, such that we are in the BOAS regime. This is the case for any finite $\mathrm{Oh}$.  

For intermediate $\alpha$, the problem is solved numerically from (\ref{nd_T}) where $\bar p$ and $\bar T$ are chosen such that the solution presents the desired wavelength $\lambda^*$, while conserving the volume. For $\mathrm{Oh}=1$, the size of the bead and the string are plotted as the orange solid lines in Fig.~\ref{fig:phase_plot}. Clearly, the predictions by the orange lines provide an excellent match to the full three-dimensional numerical solutions (green squares). This agreement shows that, indeed, the nonlinear states are selected by the dynamical evolution of the thread. 

Let us summarise the scenario that emerges, by recapitulating Fig.~\ref{fig:phase_plot}. The dynamically selected solutions appear for $\alpha > 6$, which is the threshold for the instability. Near the threshold, $\alpha \rightarrow 6^+$, the fastest growing wavelength $\lambda^*$ goes to infinity and the selected solutions naturally converge to the COAS morphology (this limit is marked by the two orange circles in Fig.~\ref{fig:phase_plot}). For all values of $\alpha$ the minimum thickness (lower orange line) closely follows the string size (lower dashed line, $\bar h_1$). However, the maximum radius (upper orange line) quickly falls below from the thickness predicted from the COAS-analysis (upper dashed line, $\bar h_2$). This means that the COAS configuration is typically observed up to $\alpha \approx 10$, in line with experiments~\cite{mora10}. For larger $\alpha$, the selected bead size converges to a value obtained by expansion of (\ref{eq:selection}),

\begin{equation}
\bar{R}=\left(2+3\sqrt{2}\mathrm{Oh}\right)^{1/6}\left( {3\pi\over 2}\right)^{1/3}.
\label{bead_asymp}
\end{equation}
The corresponding value for $\mathrm{Oh}=1$ is indicated as black dashed line in Fig.~\ref{fig:phase_plot}. Similar curves can be computed for any value of $\mathrm{Oh}$ (not shown). Each of these curves will start at the same point at $\alpha=6$, but converge to (\ref{bead_asymp}) for large $\alpha$.

Finally, we conclude this section by directly comparing the profiles obtained from the full three-dimensional numerics and the slender analysis. Figure~\ref{fig:dyn_profiles} show the profiles obtained by two methods. It may come as a surprise to the reader that the slender theory results exhibit a nearly perfect agreement with the simulation profiles, particularly in the transition region between the thin and the thick part. This is due to the fact that our expression of curvature in the capillary energy is exact, which enables the model to capture moderate to large slopes. Apparently, the approximations in the elastic energy have a small effect on the profiles. We will return to the relation between the slender formulation and the three-dimensional problem in Sec.~\ref{sec:conservation}.


\section{Analysis beyond the slender approximation: conservation laws}
\label{sec:conservation}

The reason that the slender approximation does so well in capturing the numerical observations is hidden in its conservative nature. For the stationary states, the two quantities $\bar{p}$ and $\bar{T}$ remain constant along the length of the deformed cylinder; for this reason we were able to simplify the expressions in regions where the shapes simplify to portions of a cylinder or of a sphere. In this section, we demonstrate that $\bar{p}$ and $\bar{T}$ emerge from conservation laws that are valid beyond the slender assumption, and that, at equilibrium, (\ref{eq_P},\ref{eq_T}) can be derived from the full elasto-capillary problem without any approximation.

\subsection{Translational invariance}

To express the general conservation laws of elasticity, we write the elastic energy as a functionals of the mapping $\mathbf x(\mathbf X)$, in the form

\begin{equation}\label{eq:fjacco}
\mathcal F_{\rm el}\left[\mathbf x \right]  = \int d\mathbf X \, \,W\left(\mathbf x, \frac{\partial \mathbf x}{\partial \mathbf X},\mathbf X\right).  
\end{equation}
We remind that $\mathbf F = \frac{\partial \mathbf x}{\partial \mathbf X}$ contains the elastic deformation, and in the present problem $W(\mathbf F)$ without any dependence on the current coordinate $\mathbf x$ or on the material coordinate $\mathbf X$. It is instructive, however, to retain the formalism with $W(\mathbf x, \mathbf F, \mathbf X)$. Namely, an explicit dependence on $\mathbf x$ will act as a potential energy and will give rise to external forces. An explicit dependence of $W$ on the material coordinate $\mathbf X$ implies inhomogeneities of the system in the reference state. In what follows below we will closely follow the work by Eshelby~\cite{E75}, and separately exploit the consequences of both $\partial W/\partial \mathbf x=0$ and $\partial W/\partial \mathbf X=0$. The invariance with respect to $\mathbf x$ gives a force balance, while indeed a second conservation law originates from the invariance with respect to $\mathbf X$. Both laws were derived by Eshelby~\cite{E75} as a form of Noether's theorem, which expresses that each continuous symmetry is accompanied by a conservation law. Subsequently, these conservation laws are worked out in the context of the present problem, without making use of the slender hypothesis.

\begin{figure}[t]
    \centering
    \includegraphics[width=0.85\linewidth]{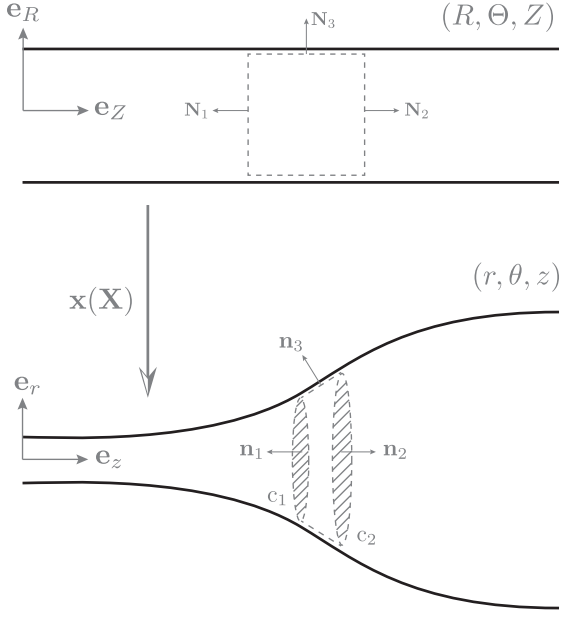}
    \caption{Conservation laws in reference (top panel) and current (bottom panel) configurations. The dashed lines represent surfaces enclosing a material volume in both of the configurations. $\mathbf{N}_i$ ($\mathbf{n}_i$) are the unit normals to the reference (current) surfaces.}
    \label{fig:mapping}
\end{figure}

\subsubsection{Current configuration}

In the absence of external force fields, the elastostatic equation is in the form of a conservation law $\textbf{div}\cdot\mathbf\sigma=0$, expressing the divergence of the Cauchy stress $\mathbf\sigma$ on the current configuration. For  incompressible elastic media the Cauchy stress reads $\mathbf\sigma=\pd{W}{\mathbf F}\cdot\mathbf{F}^T$. Here we interpret this results as a consequence of translational invariance of the energy (\ref{eq:fjacco}) with respect to the current position $\mathbf x$. Remembering that $\mathbf F = \frac{\partial \mathbf x}{\partial \mathbf X}$, the Euler-Lagrange equation for (\ref{eq:fjacco}) reads

\begin{equation}
\frac{\delta \mathcal F_{\rm el}}{\delta \mathbf x} 
= \frac{\partial W}{\partial \mathbf x} - \mathbf{Div}\left(\frac{\partial W}{\partial \mathbf F }\right)=0.
\end{equation}
Translational invariance with respect to the current coordinate implies $\partial W/\partial \mathbf x=0$, such that the equilibrium condition simplifies to a vanishing divergence on the reference domain (denoted by $\mathbf{Div}$). Transformed from the reference domain to the current domain, this gives the above-mentioned $\textbf{div}\cdot\mathbf\sigma=0$ on the current domain.


Physically, we can of course interpret this as a balance of forces on a material volume indicated by the dashed lines in the bottom panel of Fig.~\ref{fig:mapping}. When expressing the divergence as a surface integral, the force balance on the volume element reads $\int_{s_1}\mathbf\sigma\cdot\mathbf{n}_1\,\id s_1+\int_{s_2}\mathbf\sigma\cdot\mathbf{n}_2\,\id s_2+\int_{s_3}\gamma\kappa\mathbf{n}_3\,\id s_3=0$, where the $\mathbf{n}_{1,2,3}$ are the unit vectors normal to the surfaces $s_{1,2,3}$,  and we used the boundary condition that normal traction on the free surface is of capillary origin $\mathbf\sigma\cdot\mathbf{n}_3=\gamma\kappa\mathbf{n}_3$. Applying Stokes theorem we further simplify the surface integral $\int_{s_3}\gamma\kappa\mathbf{n}_3\,\id s_3$ to contour integrals on the two bounding circles, indicated by $c_{1,2}$. With this, the force balance on the volume element reduces to 

\begin{equation}
\int_{s_1}\mathbf\sigma\cdot\mathbf{e}_z\,\id s_1+\oint\limits_{c_1}\gamma\,\mathbf{t}_{1}\,\id\ell =   \int_{s_2}\mathbf\sigma\cdot\mathbf{e}_z\,\id s_2 +\oint\limits_{c_2}\gamma\,\mathbf{t}_{2}\,\id\ell  ,
\label{eq:blab}
\end{equation} 
where we also used that $\mathbf{n}_1=-\mathbf{e}_z$ and $\mathbf{n}_2=\mathbf{e}_z$, and introduced the tangential vectors $\mathbf t_{1,2}$. 
By axisymmetry, the radial component of this force balance is identically satisfied, and we are left with the axial component. Given that (\ref{eq:blab}) holds for \emph{any} choice of the cross-section, i.e. for any choice of $z_{1,2}$, we can write it in the form 

\begin{equation}
\int_s\sigma_{zz}\,\id s+\oint\limits_c\gamma\left(\mathbf{t}\cdot\mathbf{e}_z\right)\,\id\ell=T,
\label{genT}  
\end{equation} 
where $T$ is constant along the thread. This is a generalized form of eq.~\eqref{T} marking the ``conservation" of axial tension.

\subsubsection{Reference configuration}

The second conservation law can be derived using the translational invariance with respect to the reference coordinate $\mathbf X$. Namely, when the material inside an elastic medium is homogeneous in the reference state, the energy should be invariant with respect to a translation in $\mathbf X$. Under the condition that $\partial W/\partial \mathbf X=0$, one can demonstrate by direct evaluation that 

\begin{equation}
\frac{\delta \mathcal F_{\rm el}}{\delta \mathbf x} \cdot \mathbf F = -\textbf{Div}\left( \pmb{\Pi}\right),
\end{equation}
where we defined Eshelby's energy-momentum tensor~\cite{E75}

\begin{equation}
 \pmb{\Pi}=\mathbf{F}^T\cdot\pd{W}{\mathbf F}-W\mathbf{I}. 
\end{equation}
Hence, at equilibrium $\textbf{Div}\left(\pmb{\Pi}\right)=0$ emerges as the conservation law associated to invariance with respect to $\mathbf X$. 

Once again the conservation law can be written as a surface integral over a material volume shown in the top panel of fig.~\ref{fig:mapping}, where the $\mathbf{N}_{1,2,3}$ are the normals of the bounding surfaces in the reference configuration. This surface integral is the large-deformation equivalent of the $J$-integral in fracture mechanics~\cite{E75}, and is also referred to as configurational force balance -- in the absence of cracks or defects, i.e. translational invariance in $\mathbf X$, the surface integral vanishes. 

Since $\mathbf{N}_1=-\mathbf{e}_Z$, $\mathbf{N}_2=\mathbf{e}_Z$ and $\mathbf{N}_3=\mathbf{e}_R$ (cf. Fig.~\ref{fig:mapping}), the surface integral reads $-\int_{S_1}\pmb{\Pi}\cdot\mathbf{e}_Z\,\id S_1+\int_{S_2}\pmb{\Pi}\cdot\mathbf{e}_Z\,\id S_2+\int_{S_3}\pmb{\Pi}\cdot\mathbf{e}_R\,\id S_3=0$. Once again, the radial component of the surface integral is trivially satisfied by symmetry. Owing to the shear free cylindrical surface, i.e. $\mathbf{e}_Z\cdot\pmb{\Pi}\cdot\e_{R}=0$, the axial component of the configurational force balance reduces to

\begin{equation}
\int_S\Pi_{ZZ}\,\id S=k,
\label{eshelby}
\end{equation}
where $k$ is a constant, and $S$ is any cross section on the reference configuration
 space. Below we will show how (\ref{eshelby}) simplifies to the
result for the pressure $p$, previously obtained in the slender approximation.

\subsection{Neo-Hookean solid: recovering the slender results}

We now demonstrate how the slender results (\ref{eq_P},\ref{eq_T}) are recovered without any approximation -- which implies that the key features of Fig.~\ref{fig:phase_plot} are exact, and do not rely on the slender hypothesis. 
We start by noting that an axisymmetric deformation is characterized by the set of mapping $r=r(R,Z)$, $\theta=\Theta$, and $z=z(R,Z)$. For a neo-Hookean solid with elastic energy density $W={\mu\over 2}\left(\textup{Tr}(\mathbf{F}\cdot\mathbf{F}^T)-3\right)$, we can express the axial components of both the Cauchy's stress and Eshelby's stress tensor as~\cite{negahban2012}
\begin{equation}
\sigma_{zz}=\mu B_{zz}-p=\mu\left[\left(\pd{z}{Z}\right)^2+\left(\pd{z}{R}\right)^2\right]-p,
\label{}
\end{equation} and
\begin{equation}\begin{split}
\Pi_{ZZ}&=\mu C_{ZZ}-p-W={\mu\over2}\left[\left(\pd{r}{Z}\right)^2+\left(\pd{z}{Z}\right)^2\right]-p\\
&-{\mu\over 2}\left[\left(\pd{r}{R}\right)^2+\left(\pd{z}{R}\right)^2+\left({r\over R}\right)^2-3\right].
\label{}
\end{split}
\end{equation} Here $\mathbf{B} = \mathbf{F}\cdot\mathbf{F}^T$ and $\mathbf{C}=\mathbf{F}^T\cdot\mathbf{F}$ are the strain tensors in current and reference configurations respectively. In general, one is able to evaluate these quantities only after the mapping is solved. Here, however, we exploit their conservative nature and evaluate them at regions where the cylinder is locally flat. In a flat region, $\pd{r}{Z}=\pd{z}{R}=0$ and $\pd{r}{R}=h/h_0$, $\pd{z}{Z}=h_0^2/h^2$. Thus the two stresses take following forms
\begin{equation}
\sigma_{zz}=\mu{h_0^4\over h^4}-p,
\label{}
\end{equation} and
\begin{equation}
\Pi_{ZZ}={\mu\over 2}{h_0^4\over h^4}-{\mu\over 2}\left({2h^2\over h_0^2}-3\right)-p.
\label{}
\end{equation} The unknown $p$ in the equations above is evaluated using the normal stress boundary condition at the free surface, $\sigma_{rr}=\gamma\kappa=-\gamma/h$. Since $\sigma_{rr} = \mu h^2/h_0^2 -p$, we find $p=\gamma/h+ \mu h^2/h_0^2$. Now we are in a position to evaluate the eqs.~\eqref{genT} and \eqref{eshelby}. Within  a flat region \eqref{genT} simply becomes $\sigma_{zz} h^2+2\gamma h=\gamma h+\mu h_0^4/h^2-\mu h^4/h_0^2=T$, where we have absorbed the constant $\pi$ into $T$. Using the dimensionless quantities introduced in sec.~\ref{fem}, we find the expression of tension as
\begin{equation}
\bar{T}=\bar{h}+{1\over\alpha}\left({1\over\bar{h}^2}-\bar{h}^4\right).
\label{}
\end{equation} 
This is exactly the same as \eqref{eq_T}, but now derived without any slender approximation. Similarly \eqref{eshelby} becomes $-\mu h_0^4/2 h^4+2 \mu h^2/h_0^2+\gamma/h=K$ where $K=(-k/\pi h_0^2)+3\mu/2$. In dimensionless form we then find
\begin{equation}
\bar{K}={1\over\bar{h}}+{1\over\alpha}\left(2\bar{h}^2-{1\over2\bar{h}^4}\right),
\label{barK}
\end{equation} 
which is \eqref{eq_P}. 

We have thus demonstrated that (\ref{eq_P},\ref{eq_T}) are valid beyond the slender approximation. This means that the COAS features in Fig.~\ref{fig:phase_plot} (blue dashed lines) are exact. Similarly, the scaling of the thread thickness in the BOAS regime, i.e. $\bar{h}_{th}=(2\alpha)^{-1/3}$ is exact, as also already inferred in~\cite{eggers_herrada_snoeijer_2020}. Namely, for spherical beads of radius $R$, capillarity dominates over elasticity to give, $\Pi_{ZZ}=-p=-2\gamma/R$, where we use the capillary pressure of a sphere. The bead connects to a thin thread that must have the same value of $\bar K$ associated to this $\Pi_{ZZ}$. Thus evaluating~\eqref{barK} for a string, dropping the subdominant elastic term proportional to $\bar{h}_{th}$, and equating it to $\bar{K}$ for a bead, we obtain 
\begin{equation}  
{1\over\bar{h}_{th}}-{1\over2\alpha\bar{h}^4_{th}}={2\over\bar{R}}.
\label{}
\end{equation} 
Finally, if the bead size if much larger than the thread thickness ($\bar{R}\gg\bar{h}_{th}$), we recover the scaling $\bar{h}_{th}=(2\alpha)^{-1/3}$. This concludes our discussion on the conservation laws to show that the essential features of the phase map shown in Fig.~\ref{fig:phase_plot} remain valid beyond the slender approximation.

\section{Discussion}\label{sec:discussion}

In summary, we present a detailed study on the capillary driven instability of soft solid cylinders including the threshold of instability, a phase map of all possible morphologies and the dynamical selection thereof. The key result from the present work is that there is not a unique equilibrium solution for a given stiffness and surface tension, and previously derived beads-on-a-string (BOAS) and cylinders-on-a-string (COAS) configurations can actually coexist. The final pattern that emerges is selected via a dynamical process -- very much like distribution of drop sizes in the conventional ``liquid" Rayleigh-Plateau instability. While for many results we have used a slender approximation, to efficiently explore the space of solutions and to describe the dynamical selection mechanism, the results are in excellent agreement with thee-dimensional numerical simulations. In addition, we were able to derive various analytical results without invoking the slender assumption. As such, the phase map in Fig.~\ref{fig:phase_plot} offers a general description of the nonlinear states of the elastic Rayleigh-Plateau instability. 

The experiments on which cylinders have been observed are using
agar gels, which allows values of up to $\alpha\approx 10$. In this
range of values radial stretch is still significant, so it is 
difficult to see a good realisation of the BOAS structure, characterized
by almost spherical beads. By contrast in polymer solutions, $\alpha$
can easily reach values of about 200, leading to much more pronounced
BOAS structures \cite{CEFLM06}. Once a first balance of surface tension and
elasticity is reached, elastic stresses relax exponentially in time,
with a corresponding thinning of the thread, which can be followed
by at least another order of magnitude \cite{DHEB20}. As shown
by \cite{eggers_herrada_snoeijer_2020}, the resulting shape during
this relaxation phase is strictly identical to the purely elastic BOAS structure.

Our predictions on the selection of COAS versus BOAS are thus in good agreement with experimental observations. Future work could be dedicated to extending the dynamical evolution beyond the model presented here, which is essentially based on an Oldroyd-B fluid in the limit of infinite relaxation time. Similarly, one can envisage an elasticity beyond the neo-Hookean solid, for example to account for the finite extensibility inside the thin threads.


{\bf Acknowledgments.} 

M.K. and J.H.S. acknowledges support from NWO through VICI Grant No. 680-47-632,
and A.P. from European Research Council (ERC) Consolidator Grant No. 616918.
J. E. acknowledges the support of Leverhulme Trust International Academic
Fellowship IAF-2017-010.
M.A.H thanks the Ministerio de Economia y Competitividad and the Junta de Andalucia for partial support under the Projects No. PID2019-108278-RB-C31 and No. PAIDI: P18-FR-3623 respectively.

\appendix

\section{Numerics via mapping technique}\label{app:mapping}

The numerical technique used in this study is a variation of that
developed in \cite{Herrada2016} for interfacial flows and extended in \cite{eggers_herrada_snoeijer_2020} for viscoelastic fluids. The spatial physical domain  is mapped onto a rectangular domain by means of a non singular mapping

$$
r=f(\xi,\chi,t),\quad z=g(\xi,\chi,t),$$
$$R=F(\xi,\chi,t),\quad Z=G(\xi,\chi,t),\quad [0\leq \xi\leq L]\times [0\leq \chi\leq 1],
$$
where functions $f$, $g$, $F$ and $G$ should be obtained as a part of the solution. To determine these functions, the following equations have been used
\begin{eqnarray}
 g &=&\xi, \label{mappingz} \\
  F &=& h_0.\label{rigid}
\end{eqnarray}
Equation (\ref{mappingz}) guarantees that the same discretization  used for the $\xi$ variable is automatically applied to variable $z$. Finally, equation (\ref{rigid}) indicates that at the initial stage  the interface is a perfect cylinder of radius $h_0$.

Some additional boundary conditions  for the shape functions are needed to close the problem.  At $z =\xi=0$
\begin{equation}
g =\frac{\partial f}{\partial \chi}=0,
\end{equation}
at  $z =\xi= L$,
\begin{equation}
g=L,\quad f=h_0,
\end{equation}
and at  $r =\chi=0$,
\begin{equation}
g=\xi,\quad f=0.
\end{equation}

The unknown variables in the numerical domain are 
$f$, $g$, $p$, $v_{r}$, $v_{z}$, $F$ and $G$. 
All the derivatives appearing in the governing equations are  expressed  in term of $\chi$, $\xi$ and $t$. Then, the resulting equations are discretized in the $\chi$  direction with $n_{\chi}$  Chebyshev spectral collocation points. In the $\xi$ direction, we use fourth-order finite differences with $n_{\xi}$  equally spaced points. The results presented in this work were carried out using $n_{\xi_1}=5500$ and $n_{\chi}=10$.

To compute the unsteady 2D solution, all the equations are solved together (monolithic scheme)  using a Newton–Raphson technique. In this case, second order backward differences are used to compute the time with a variable time step. One of the main characteristics of this procedure is that the elements of the Jacobian $\mathcal{J}^{(p,q)}$ of the discretized system of equations  $\mathcal{J}^{(p,q)}\Delta \Psi^{(q)}=-\mathcal{F}^{(p)}$  for the base solution updates $\Delta \Psi^{(q)}$ ($q=1,2,...,n\times N$ stands for the values of the n unknowns at the N grid points) are
obtained by combining analytical functions and collocation matrices. This allows taking advantage of the sparsity of the resulting matrix to reduce the computational time on each Newton step.

As explained by \cite{Herrada2016}, the  numerical procedure can be also easily adapted to solve the eigenvalue
problem which determines the linear stability of the 1D basic solution, of a uniform thread with vanishing velocity and vanishing elastic stress. In this case, the temporal and axial derivatives are computed assuming the dependence  

\begin{equation}
\Psi(z,r;t)=\Psi_b(r)+\epsilon \delta \Psi(r)e^{-i\omega t+i qz} \quad (\epsilon\ll 1). \label{perturbations}
\end{equation}
Here, $\Psi(z, r; t)$ represents any dependent variable, $\Psi_b( r)$ and $\delta \Psi(r)$ stand for
the 1D base (steady) solution and the radial (1D) dependence of the eigenmode, respectively,
while $\omega=\omega_r+i\omega_i$ is the eigenfrequency and $q$ (real) is the axial wavenumber. The radial dependence of the linear perturbation $\delta \Psi^{(q)}$, for a given axial wavenumber $q$, is the solution to the generalised eigenvalue problem $\mathcal{J}^{(p,q)}_b\delta\Psi^{(q)}=i\omega\mathcal{Q}^{(p)}_b \delta \Psi^{(q)}$, where $\mathcal{J}^{(p,q)}_b$ is the Jacobian of
the system evaluated with the basic solution $\Psi_b^{(q)}$, and with all axial derivatives computed according to (\ref{perturbations}), while
$\mathcal{Q}^{(p,q)}_b$ accounts for the temporal
dependence of the problem. This  generalised eigenvalue problem is solved using MATLAB EIG function.





%

\end{document}